\documentclass[iop]{emulateapj}
\usepackage{apjfonts}
\usepackage{graphicx}

\def\ltsima{$\; \buildrel < \over \sim \;$}
\def\simlt{\lower.5ex\hbox{\ltsima}}
\def\gtsima{$\; \buildrel > \over \sim \;$}
\def\simgt{\lower.5ex\hbox{\gtsima}}

\usepackage{color}

\shorttitle{Energy feedback of XRBs at high $z$}
\shortauthors{T. Fragos}

\begin{document}

\title{Energy Feedback from X-ray Binaries in the Early Universe}

\author{T.\ Fragos$^{1,2}$,  B.\ D.\ Lehmer$^{3,4}$, S. Naoz$^{2,\dagger}$, A.\ Zezas$^{5,6,1}$, A.\ Basu-Zych$^{4}$}
\altaffiltext{1}{Harvard-Smithsonian Center for Astrophysics, 60 Garden Street, Cambridge, MA 02138 USA}
\altaffiltext{2}{Institute for Theory and Computation, Harvard-Smithsonian Center for Astrophysics, 60 Garden Street, Cambridge, MA 02138 USA}
\altaffiltext{3}{The Johns Hopkins University, Homewood Campus, Baltimore, MD 21218, USA}
\altaffiltext{4}{NASA Goddard Space Flight Centre, Code 662, Greenbelt, MD 20771, USA}
\altaffiltext{5}{Department of Physics, University of Crete, P.O. Box 2208, 71003 Heraklion, Crete, Greece}
\altaffiltext{6}{IESL, Foundation for Research and Technology, 71110 Heraklion, Crete, Greece}
\altaffiltext{$\dagger$}{Einstein Fellow}
\email{tfragos@cfa.harvard.edu}

\begin{abstract}

X-ray photons, because of their long mean-free paths, can easily escape the galactic environments where they are produced, and interact at long distances with the inter-galactic medium, potentially having a significant contribution to the heating and reionization of the early Universe. The two most important sources of X-ray photons in the Universe are active galactic nuclei (AGN) and X-ray binaries (XRBs). In this Letter we use results from detailed, large scale population synthesis simulations to study the energy feedback of XRBs, from the first galaxies ($z\sim 20$) until today. We estimate that X-ray emission from XRBs dominates over AGN at $z\gtrsim 6-8$. The shape of the spectral energy distribution of the emission from XRBs shows little change with redshift, in contrast to its normalization which evolves by $\sim 4$ orders of magnitude, primarily due to the evolution of the cosmic star-formation rate. However, the metallicity and the mean stellar age of a given XRB population affect significantly its X-ray output. Specifically, the X-ray luminosity from high-mass XRBs per unit of star-formation rate varies an order of magnitude going from solar metallicity to less than $10\%$ solar, and the X-ray luminosity from low-mass XRBs per unit of stellar mass peaks at an age of $\sim 300\,\rm Myr$ and then decreases gradually at later times, showing little variation for mean stellar ages $\gtrsim 3\,\rm Gyr$. Finally, we provide analytical and tabulated prescriptions for the energy output of XRBs, that can be directly incorporated in cosmological simulations.

\end{abstract}


\maketitle

\section{INTRODUCTION}

An important cosmic milestone is the appearance of the first luminous objects, which ends the era of the cosmic dark ages and begins the era of heating and reionization of the intergalactic medium (IGM). 
The energy and momentum output from stars is believed to be a major feedback mechanism, along with the feedback from active galactic nuclei (AGN), that regulate galaxy formation and the re-ionization epoch \citep[e.g.][]{SMH2005,Sales2010,Hopkins2011,Dib2011,FGQH2013}. To date, the vast majority of cosmological simulation studies consider only the feedback from massive O stars and AGN via their ionizing ultra-violet (UV) radiation and deposition of mechanical energy and momentum in the vicinities of star-forming regions and accreting supermassive black holes \citep[BH; e.g.,][]{McQ2007,Stinson2013,SM2013}.

 UV photons are easily absorbed by neutral hydrogen, and they are efficient in ionizing it. Since the energy output of massive stars and AGN peaks at UV wavelengths the radiation from the first galaxies is expected to eventually ionize the neighboring IGM. The Swiss-cheese paradigm, in which the neutral background gas is spotted with spherical regions of hot and ionized gas by starburst galaxies and quasars \citep[e.g.][]{McQ2007b}, is widely accepted \citep[e.g.,][]{Loeb2001}.

On the contrary, the more energetic X-ray photons, because of their long mean-free paths, can escape the galactic environments where they are produced, and interact at Mpc scales with the IGM. This could potentially result in a smoother spatial distribution of ionized regions \citep[e.g.][]{OG1996,PF2007}, and perhaps more importantly, in an overall warmer IGM \citep[e.g.,][]{Mirabel2011,Zoltan2011}. The two most important sources of X-ray photons in the Universe are AGN and X-ray binaries (XRBs). Current constraints show that AGN provide at least an order-of-magnitude higher X-ray luminosity per unit volume over XRBs from $z \approx$~0--3 \citep[e.g. ][]{BasuZych2013}. These constraints indicate that the X-ray luminosity per unit volume for XRBs begins to ``catch up'' to that of AGN going to the highest measurable redshifts (e.g., $z \approx$~3--4). Therefore, it is plausible that at even higher redshifts ($z \simgt$~4--5), the X-ray emission from XRBs becomes important or even dominant.

The possible feedback processes from XRBs at high redshifts ($z\gtrsim 6$) has been the topic of several papers in the last few years. \citet{Mirabel2011} considered primordial population-III binaries  ($z\geq 6$), and proposed that besides the ultraviolet radiation from massive stars, feedback from accreting BHs in high-mass XRBs (HMXBs) was an additional, important source of heating and re-ionization of the IGM. 
\citet{JS2012} studied how XRBs inject energy in their local environments before luminous supernovae contribute significantly to feedback. They argue that XRBs can also assist in keeping gas hot long after the last core-collapse SN has exploded. 
\citet{Power2012} explored the ionizing power of HMXBs at high redshifts using simple Monte Carlo modeling for their formation and the Galactic HMXB Cygnus X-1 as a spectral template for their emission in X-rays. 


In this Letter, we go for the first time beyond the order of magnitude calculations or simple rate models for the evolution of XRBs on cosmological timescales. We use results from a detailed large scale population synthesis simulation of the evolution of XRBs across cosmic time \citep{Fragos2013}, which has been calibrated to all available observations of XRB populations in the local and low redshift Universe ($z\lesssim 4$), to study the energy output of XRBs at high redshift ($z\gtrsim 6-8$). We compare our model with the X-ray emission of AGN at the same redshifts and derive prescriptions for the feedback of XRBs that can be incorporated in future cosmological simulations.

\section{X-ray Binary Population Synthesis Models}

\begin{figure*}
\centering
\plottwo{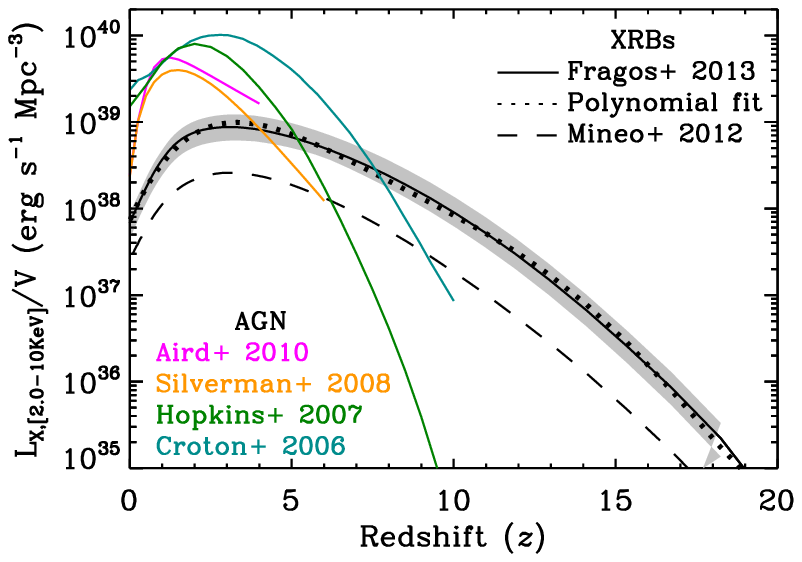}{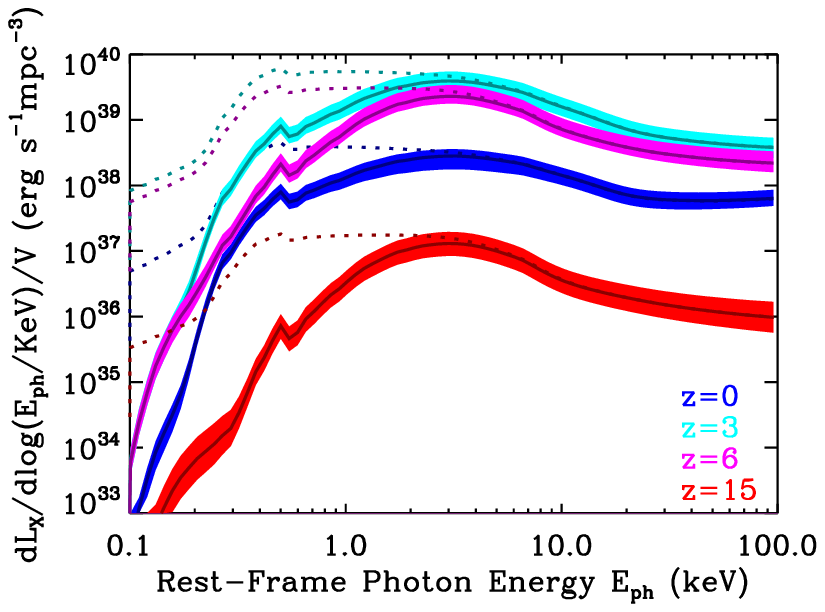}
\caption{\label{LxV} \textbf{(Left panel)} X-ray luminosity per unit co-moving volume in the $2-10\,\rm KeV$ band as a function of redshift. The grey shaded area shows the differences between the predictions of the six highest likelihood models by \citet{Fragos2013} for the X-ray emission coming from XRBs. The black solid line corresponds to the mean value, and the dotted dark-grey line to the polynomial fit on the mean (equation~\ref{eq_fit1}). The dashed line is derived by convolving the locally measured value of $L_{\rm X}$/SFR \citep{Mineo2012a} with the SFH of the Universe. For comparison the X-ray luminosity density of AGN in the same energy range is plotted, as reported by different AGN models and observations \citep{Croton2006,HRH2007, Silverman2008, Aird2010}. \textbf{(Right panel)} The SED of the global XRB population at four different redshifts is shown. The solid lines correspond to the mean value of the different models and the shaded area denotes the model uncertainties, assuming in both cases that the interstellar absorption is similar to the Milk Way at all redshifts. With dotted lines we show an estimate of the intrinsic (unabsorbed) SED.}
\end{figure*}

\begin{deluxetable*}{cccccccc} 
\centering
\tablecolumns{8}
\tabletypesize{\scriptsize}
\tablewidth{0pt}
\tablecaption{Synthetic SED data at different redshifts ($F(E_{ph})=dL_X/dlog(E_{ph}/KeV)/V\,\rm [erg/s/mpc^3]$). The full table is available as online-only supplemental material. \label{SEDdata}} 
\tablehead{\multicolumn{8}{c}{SED that includes interstellar absorption}
}
\startdata
\colhead{$E_{ph}$} & 
\colhead{$F(E_{ph})$ @ z=19.92} & 
\colhead{$F(E_{ph})$ @ z=18.24} & 
\colhead{$F(E_{ph})$ @ z=16.72} & 
\colhead{$F(E_{ph})$ @ z=15.34} & 
\colhead{$F(E_{ph})$ @ z=14.09} & 
\colhead{$F(E_{ph})$ @ z=12.94} & 
\colhead{$\cdots$} \\
\hline

1.020E+00      &    4.590E+34      &      2.969E+35      &      1.139E+36       &     3.455E+36       &     9.036E+36      &      2.034E+37      &    $\cdots$ \\   
1.115E+00      &    5.875E+34      &      3.799E+35      &      1.458E+36       &     4.419E+36       &     1.155E+37      &      2.598E+37      &    $\cdots$ \\   
1.219E+00      &    7.256E+34      &      4.692E+35      &      1.800E+36       &     5.455E+36       &     1.425E+37      &      3.204E+37      &    $\cdots$ \\
1.333E+00      &    8.491E+34      &      5.490E+35      &      2.106E+36       &     6.380E+36       &     1.667E+37      &      3.745E+37      &    $\cdots$ \\
1.457E+00      &    9.874E+34      &      6.383E+35      &      2.448E+36       &     7.417E+36       &     1.937E+37      &      4.351E+37      &    $\cdots$ \\
1.593E+00      &    1.139E+35      &      7.364E+35      &      2.824E+36       &     8.555E+36       &     2.234E+37      &      5.016E+37      &    $\cdots$ \\
1.741E+00      &    1.283E+35      &      8.291E+35      &      3.179E+36       &     9.630E+36       &     2.514E+37      &      5.645E+37      &    $\cdots$ \\
1.903E+00      &    1.377E+35      &      8.898E+35      &      3.412E+36       &     1.033E+37       &     2.698E+37      &      6.056E+37      &    $\cdots$ \\
2.080E+00      &    1.495E+35      &      9.665E+35      &      3.706E+36       &     1.122E+37       &     2.930E+37      &      6.576E+37      &    $\cdots$ \\
$\vdots$ & $\vdots$ & $\vdots$ & $\vdots$ & $\vdots$ & $\vdots$ & $\vdots$ & $\ddots$ \\

\hline \hline
\multicolumn{8}{c}{Intrinsic (unabsorbed) SED} \\
\hline
\colhead{$E_{ph}$} & 
\colhead{$F(E_{ph})$ @ z=19.92} & 
\colhead{$F(E_{ph})$ @ z=18.24} & 
\colhead{$F(E_{ph})$ @ z=16.72} & 
\colhead{$F(E_{ph})$ @ z=15.34} & 
\colhead{$F(E_{ph})$ @ z=14.09} & 
\colhead{$F(E_{ph})$ @ z=12.94} & 
\colhead{$\cdots$}\\
\hline
1.020E+00       &     2.288E+35       &     1.478E+36       &     5.666E+36      &      1.716E+37       &     4.476E+37      &      1.004E+38    &    $\cdots$ \\    
1.115E+00       &     2.323E+35       &     1.500E+36       &     5.753E+36      &      1.742E+37       &     4.545E+37      &      1.020E+38    &    $\cdots$ \\    
1.219E+00       &     2.331E+35       &     1.505E+36       &     5.773E+36      &      1.748E+37       &     4.559E+37      &      1.023E+38    &    $\cdots$ \\    
1.333E+00       &     2.339E+35       &     1.510E+36       &     5.791E+36      &      1.753E+37       &     4.573E+37      &      1.026E+38    &    $\cdots$ \\    
1.457E+00       &     2.331E+35       &     1.505E+36       &     5.771E+36      &      1.747E+37       &     4.557E+37      &      1.022E+38    &    $\cdots$ \\    
1.593E+00       &     2.339E+35       &     1.510E+36       &     5.790E+36      &      1.753E+37       &     4.572E+37      &      1.026E+38    &    $\cdots$ \\    
1.741E+00       &     2.332E+35       &     1.506E+36       &     5.772E+36      &      1.747E+37       &     4.558E+37      &      1.023E+38    &    $\cdots$ \\    
1.903E+00       &     2.319E+35       &     1.497E+36       &     5.740E+36      &      1.738E+37       &     4.533E+37      &      1.017E+38    &    $\cdots$ \\    
2.080E+00       &     2.305E+35       &     1.488E+36       &     5.706E+36      &      1.727E+37       &     4.505E+37      &      1.011E+38    &    $\cdots$ \\    
$\vdots$ & $\vdots$ & $\vdots$ & $\vdots$ & $\vdots$ & $\vdots$ & $\vdots$ & $\ddots$ \\
\enddata
\end{deluxetable*}

Using the {\tt StarTrack} population synthesis code \citep{Belczynski2002, Belczynski2008}, \citet{Fragos2013} performed a large scale population synthesis study that models the X-ray binary populations from the first galaxies of the Universe until today. They used as input to their modeling the Millennium II Cosmological Simulation \citep{Boylan2009} and the updated semi-analytic galaxy catalog by \citet{Guo2011} to self-consistently account for the star-formation history (SFH) and metallicity evolution of the Universe. Their models, which were constrained by the observed X-ray properties of local galaxies \citep{TG2008, Lehmer2010, Boroson2011, Mineo2012a}, gave predictions about the global scaling of emission from X-ray binary populations with properties such as SFR and stellar mass, and the evolution of these relations with redshift. Although these models were only constrained to observations of the local Universe, they have been shown to be in excellent agreement with X-ray observations of high redshift normal galaxies \citep{Mineo2012b,Tremmel2013,BasuZych2013,BasuZych2013b}.

In this work, we adopt the six highest likelihood models by \citet{Fragos2013} which were also the six models which satisfy within one standard deviation all the observational constraints simultaneously \citep[see discussion in Section 4.2 of ][]{Fragos2013}. Instead of just choosing the maximum likelihood model, we assume that the differences among these six models represent in some sense the uncertainty in the model's predictions with regard to the redshift evolution of global scaling relations of emission from XRB populations.

We should not here that the population synthesis models we use assume an initial mass function that does not evolve with redshift. However, there are theoretical arguments and indirect observational evidence suggest that the stellar IMF may evolve with time, becoming flatter at higher redshift \citep[e.g.][]{Dokkum2008}. \citet{Fragos2013} showed that a flatter initial mass function results to more luminous XRB populations. Furthermore, our models are applicable to stellar population with metallicities down to $Z=10^{-4}$, and  do not take into account the first generation of metal-free POP-III star, which can potentially have significantly different evolution compared to enriched populations.

\section{Energy Output of the X-ray binary population}

The emission of X-ray photons in the local Universe is dominated by AGN, whose X-ray flux is approximately an order of magnitude stronger than that coming from XRB populations of normal galaxies. This picture is gradually changing as we move to higher redshifts, with the X-ray luminosity density coming from normal galaxies increasing faster than than from AGN. The left panel of Figure~\ref{LxV} shows the redshift evolution of the X-ray luminosity density (X-ray luminosity per unit co-moving volume $L_{\rm X}/V$) coming from XRBs, as predicted by the models of \citet{Fragos2013}. On the same figure, several models and observational estimates for the X-ray luminosity density of AGN, as a function of redshift, are shown \citep{Croton2006,HRH2007, Silverman2008, Aird2010}. It is evident that at $z\gtrsim 6-8$, XRBs dominate the X-ray luminosity density, since the massive BHs at the centers of galaxies did not have enough time yet to grow and outshine the XRBs.

For comparison, we also show the contribution from XRBs that one would get if the locally measured value of the X-ray luminosity from HMXBs per unit of SFR \citep[$L_{\rm X}$/SFR;][]{Mineo2012a} is convolved with the SFH of the Universe. The adopted SFH comes from the Millenium II simulation \citep{Boylan2009, Guo2011}. We see that the contribution of XRBs is underestimated both at low and at high redshift. At low redshift, the contribution from low-mass XRBs (LMXB) is neglected as the measurement by \citet{Mineo2012a} was focused on the HMXB emission of starburst galaxies. We should remind here that in the local Universe it is the LMXBs that are dominating the XRB luminosity density of the Universe, with HMXBs starting to dominate only at $z\gtrsim 2.5$ \citep{Fragos2013}. At high redshift, a simple convolution of the locally measured $L_{\rm X}$/SFR with the SFH neglects the effects of metallicity evolution on the stellar population. As we will show below, the X-ray luminosity from HMXBs per unit SFR varies by approximately an order of magnitude with metallicity.

\subsection{Evolution of the spectral energy distribution with redshift}

Our population synthesis models keep track of the mass-transfer rate as a function of time for every modeled XRB. From this mass-transfer rate, the bolometric luminosity is derived based on the prescriptions presented by \citet{Fragos2008,Fragos2009}, which also account for transient behavior of XRBs. However, there is no spectral information in these models. As described in detail in Section~3 of \citet{Fragos2013}, we use two samples of RXTE observations of Galactic neutron star and BH XRBs at different spectral states \citep{McCR2006,Wu2010}, for which the best-fit parameters of simple spectral models are calculated. Using these best fit spectral models, and assuming that the interstellar absorption in high redshift galaxies is similar to that in the Milky Way today, we calculate for each energy band the mean and the variance of the bolometric correction for the high-soft and low-hard states, and BH and NS XRBs separately. Combining these bolometric corrections with our population synthesis models allows us to estimate the spectral energy distribution (SED) of a XRB population. In addition, we estimate the intrinsic XRB SED (without interstellar absorption) by removing the absorption component of the observed spectral models while making the assumption that the power-law component in the high and very high state does not extend to energies below $\sim 1\,\rm KeV$, and in the low-hard state below $\sim 0.2\,\rm KeV$ \citep{SZ2006}. This allows the reader to apply a posteriori a more complicated model for interstellar absorption that can evolve with redshift.

\begin{figure*}
\centering 
\plottwo{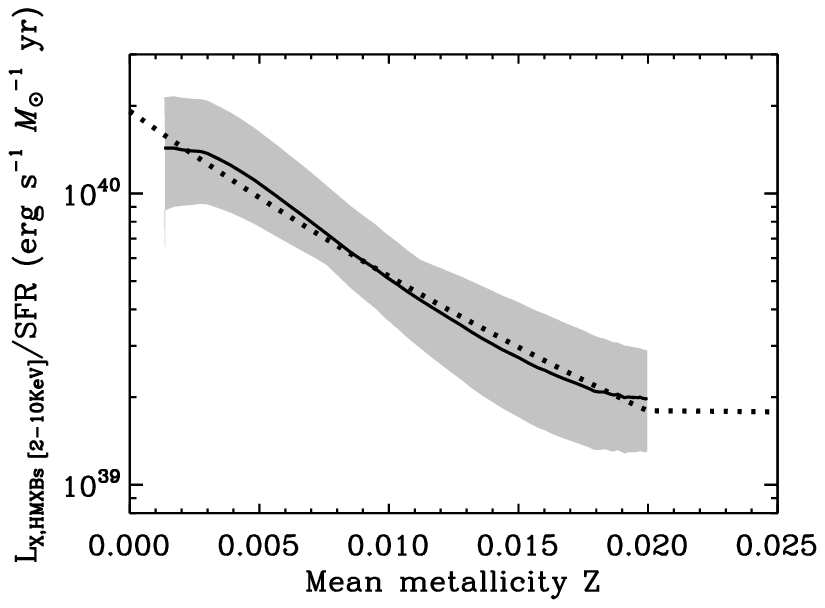}{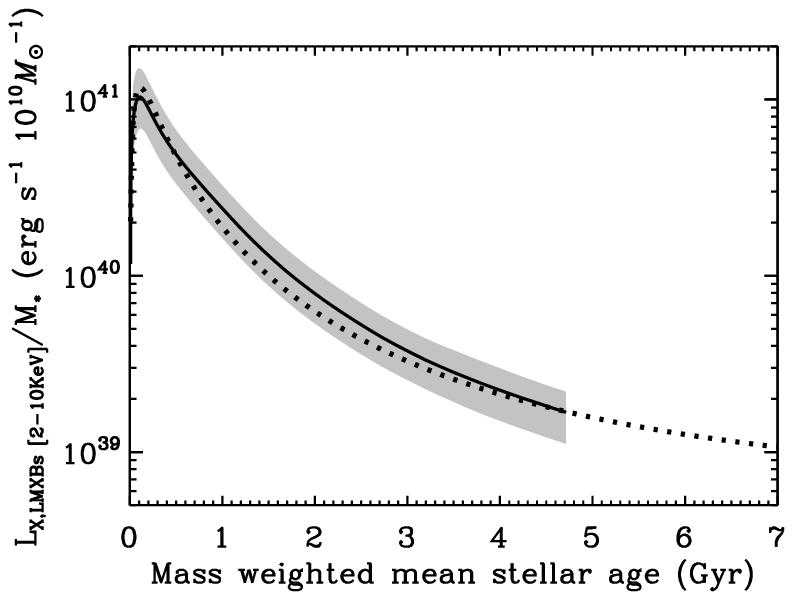} 
\caption{\label{LxSFR_LxM} X-ray luminosity in the $2-10\,\rm KeV$ band of the HMXB population, per unit SFR, as a function of the mean metallicity $Z$ of the newly formed stars (left panel) and X-ray luminosity in the $2-10\,\rm KeV$ band of the LMXB population, per unit stellar mass, as a function of the mass-weighted mean stellar age of stellar population. The grey shaded area shows the differences between the predictions of the six highest likelihood models by \citet{Fragos2013}, the solid lines corresponds to the mean value of the different models, and the dotted lines show the polynomial fits on the mean, described in equations (\ref{eq_fit2a}) and (\ref{eq_fit2b}).  } 
\end{figure*}

The right panel of Figure~\ref{LxV} shows the SED per unit co-moving volume of the global XRB population at four different redshift values. We should note here that this SED corresponds to the radiation that escapes the galaxy where the photons are produced, as the interstellar absorption is already taken into account. These photons that escape the galaxy can then interact with the IGM. We find that the shape of the SED remains approximately constant with redshift, and it is only its normalization that evolves by $\sim 4$ orders of magnitude (see left panel of Figure~1 and discussion above). The approximately constant SED shape is due to the fact that at all redshifts it is only the brightest BH XRBs in high states that dominate the integrated spectra. This is something to be expected from the shape of the X-ray luminosity functions of observed XRB populations in nearby galaxies. The shape of the X-ray luminosity functions of both LMXBs and HMXBs can be approximated by single power laws, which have exponents less than 2 \citep[$\frac{dN}{dL_{\rm X}}\propto L_{\rm X}^{-n}$, with $n<2$;][and references therein]{Fabbiano2006}. Hence, the integrated luminosity of the whole population is always dominated by the few brightest sources that are usually BH XRBs in the high-soft state. The SED data, both including interstellar absorption and intrinsic, at different redshift values  can be found in Table~\ref{SEDdata}.

Our model of the evolving X-ray luminosity density and mean X-ray SED can be used to estimate the contribution that XRBs in galaxies from $z =$~0--20 provide to the cosmic X-ray background. We integrated these models following: %
\begin{equation}
S_{\rm tot} = \frac{\Delta \Omega}{4 \pi}\frac{c}{H_0} \int_0^{20}
\frac{\rho_{\rm X}(z) K(z) dz}{(1+z)^2 \varepsilon(z)}, 
\end{equation}
where $\Delta \Omega = 3.0 \times 10^{-4}$~sr~deg$^{-2}$, $\varepsilon(z) = \sqrt{\Omega_m(1+z)^3 + \Omega_\Lambda}$, $\rho_{\rm X}$ is the X-ray luminosity density from our model (see left panel of Fig.~1), and $K(z)$ provides the redshift-dependent $K$-correction to the appropriate observed-frame energy range. The $K$-corrections are based on the SEDs shown in the right panel of Figure~1. Applying this integration to the observed-frame 0.5--2~keV bandpass, we find that XRB emission from all $z =$~0--20 galaxies is expected to contribute $S_{\rm 0.5-2~keV} \approx 2.2 \times 10^{-12}$~erg~cm$^{-2}$~s$^{-1}$~deg$^{-2}$ to the cosmic X-ray background intensity. By comparison, \citet{Lehmer2012} found that X-ray detected normal galaxies in the 4~Ms {\it Chandra} Deep Field-South contribute $2.4 \times10^{-13}$~erg~cm$^{-2}$~s$^{-1}$~deg$^{-2}$ to the resolved 0.5--2~keV background, and an additional $2.0 \times10^{-12}$~erg~cm$^{-2}$~s$^{-1}$~deg$^{-2}$ of the background remains unresolved. Our model is therefore within the allowed limits of the cosmic X-ray background \citep[also in agreement with constraints derived by][]{Dijkstra2012}. Our analysis indicates that XRBs in normal galaxies can account for up to $\sim 98\%$ of the remaining unresolved 0.5--2~keV emission. This estimate is only an upper limit as we did not take into account the absorption of X-ray photons from the IGM. In practice, we expect the contribution of XRBs to the observed cosmic X-ray background to be significantly lower.

\subsection{Prescriptions for the energy feedback of LMXB and HMXB populations.}

Our population synthesis modeling uses as input the SFH and metallicity evolution predicted by the Millennium II simulation. As a consequence, our results depend on the cosmological model and the galaxy formation and evolution prescriptions used by \citet{Boylan2009} and \citet{Guo2011}, respectively. In order to alleviate this caveat, we extracted from our models the dependence of the X-ray luminosity of HMXBs per unit SFR ($L_{X,HMXBs}$/SFR) on the mean metallicity ($Z$) of the newly formed stars and the dependence of the X-ray luminosity coming from LMXBs per unit of stellar mass ($L_{X,LMXBs}/M_{*}$) on the mass-weighted mean stellar age ($T$) of the population. Figure~\ref{LxSFR_LxM} shows that $L_{X,HMXBs}$/SFR varies by an order of magnitude going from solar metallicity to less than 10\% solar. This also indicates that at the era when HMXBs were dominating the X-ray radiation of the Universe $L_{X,HMXBs}$/SFR was approximately an order of magnitude higher than what is measured in the local Universe (at $z\gtrsim8$, $Z\lesssim 20\%Z_{\odot}$). The variation of $L_{X,LMXBs}/M_{*}$ with the mean stellar age of the population is even stronger, peaking early on at stellar population ages of $\sim 300\,\rm Myr$ and then gradually decreasing to the values observed in the local Universe. We should note here that the dependence of $L_{X,LMXBs}/M_{*}$ with stellar age, for ages above $\sim 3\,\rm Gyr$, becomes very weak, with an expected variation between ages of $\sim 3\,\rm Gyr$ and $\sim 13\,\rm Gyr$ of only a factor of a few \citep[for related observational evidence see][]{Boroson2011,ZGA2012}.

Equations~\ref{eq_fit1}-\ref{eq_fit2b} are polynomial parameterizations to our model predictions for the evolution of the X-ray luminosity density of all XRBs as a function of redshift, the dependence of the X-ray luminosity coming from HMXBs per unit SFR on the mean metallicity ($Z$) of the newly formed stars, and the dependence of the X-ray luminosity of LMXBs per unit of stellar mass on the mass-weighted mean stellar age of the population ($T$) respectively. These polynomial parameterizations are shown in Figures 1 and 2 as dotted lines.

\begin{equation} \log(L_X/V) = \alpha_0 + \alpha_1 z + \alpha_2 z^2 + \alpha_3 z^3 + \alpha_4 z^4 + \alpha_5 z^5 \label{eq_fit1} \end{equation}
$${\rm (erg\,s^{-1}\,pc^{-3})},\ where\ 0\geq z\geq 20$$

\begin{equation} \log(L_X/SFR) = \beta_0 + \beta_1 Z + \beta_2 Z^2 + \beta_3 Z^3 + \beta_4 Z^4 \, \label{eq_fit2a} \end{equation}
$$ {\rm (erg\,s^{-1}\,M_{\odot}^{-1}\, yr)},\ where\ 0\geq Z\geq 0.025$$

\begin{equation} \log(L_X/M_*) = \gamma_0 + \gamma_1 \log( T/{\rm Gyr}) + \gamma_2 \log(T/{\rm Gyr})^2 + \label{eq_fit2b} \end{equation} 
	$$ + \gamma_3 \log(T/{\rm Gyr})^3 + \gamma_4 \log(T/{\rm Gyr})^4 \,\rm (erg\,s^{-1}\,10^{10}M_{\odot}^{-1})$$
$$where\ 0\geq T\geq 13.7\,\rm Gyr$$

The parameters $\alpha_{i}\, (i=0,5)$, $\beta_{j}\, (j=0,4)$, and $\gamma_{k}\, (k=0,4)$ are derived for several widely used energy bands, both before and after taking into account interstellar absorption. The different parameter values are provided in Table~\ref{fit_parameters}.

\begin{deluxetable*}{cccccccc} 
\centering
\tablecolumns{8}
\tabletypesize{\scriptsize}
\tablewidth{0pt}
\tablecaption{List of best-fit parameter values for equations (\ref{eq_fit1}), (\ref{eq_fit2a}), and (\ref{eq_fit2b}), corresponding to different energy bands.  
\label{fit_parameters}} 
\tablehead{\multicolumn{8}{c}{Based on SED that includes interstellar absorption} 
	}
\startdata
\colhead{Energy band} & 
\colhead{$\alpha_0$} & 
\colhead{$\alpha_1$} & 
\colhead{$\alpha_2$} & 
\colhead{$\alpha_3$} & 
\colhead{$\alpha_4$} & 
\colhead{$\alpha_5$} & 
\colhead{$\chi^2/N$}\\
\hline

$0.3-8.0\rm\, keV$ 	 & 37.84 $\pm$  0.01 &  0.85 $\pm$  0.02 & -0.213 $\pm$  0.008 &  0.0212 $\pm$  0.0012 & -0.00101 $\pm$  0.00007 &  0.000018 $\pm$  0.000002 &  0.09 \\
$0.5-2.0\rm\, keV$ 	 & 37.01 $\pm$  0.01 &  0.86 $\pm$  0.02 & -0.221 $\pm$  0.008 &  0.0221 $\pm$  0.0012 & -0.00106 $\pm$  0.00007 &  0.000019 $\pm$  0.000002 &  0.10 \\
$2.0-10.0\rm\, keV$	 & 37.86 $\pm$  0.01 &  0.85 $\pm$  0.02 & -0.213 $\pm$  0.008 &  0.0212 $\pm$  0.0012 & -0.00101 $\pm$  0.00007 &  0.000018 $\pm$  0.000002 &  0.09 \\
$10-100\rm\, keV$  	 & 38.39 $\pm$  0.02 &  0.73 $\pm$  0.02 & -0.188 $\pm$  0.010 &  0.0188 $\pm$  0.0014 & -0.00091 $\pm$  0.00009 &  0.000016 $\pm$  0.000002 &  0.14 \\

\hline 
\colhead{} & 
\colhead{$\beta_0$} & 
\colhead{$\beta_1$} & 
\colhead{$\beta_2$} & 
\colhead{$\beta_3$} & 
\colhead{$\beta_4$} & 
\colhead{} & 
\colhead{} \\ 
\hline 

$0.3-8.0\rm\, keV$ 	   &    40.28 $\pm$     0.02 &   -62.19 $\pm$     1.32 &   570.07 $\pm$    13.71 & -1835.81 $\pm$    52.14 &  1970.48 $\pm$    66.27 & &  0.1751 \\
$0.5-2.0\rm\, keV$ 	   &    39.38 $\pm$     0.02 &   -61.68 $\pm$     1.31 &   565.42 $\pm$    13.64 & -1820.96 $\pm$    51.88 &  1954.61 $\pm$    65.94 & &  0.1734 \\
$2.0-10.0\rm\, keV$	   &    40.28 $\pm$     0.02 &   -62.12 $\pm$     1.32 &   569.44 $\pm$    13.71 & -1833.80 $\pm$    52.14 &  1968.33 $\pm$    66.27 & &  0.1751 \\
$10-100\rm\, keV$  	   &    40.54 $\pm$     0.02 &   -61.48 $\pm$     1.35 &   563.60 $\pm$    13.99 & -1814.95 $\pm$    53.20 &  1948.02 $\pm$    67.61 & &  0.1823 \\

\hline 
\colhead{} & 
\colhead{$\gamma_0$} & 
\colhead{$\gamma_1$} & 
\colhead{$\gamma_2$} & 
\colhead{$\gamma_3$} & 
\colhead{$\gamma_4$} & 
\colhead{} & 
\colhead{} \\ 
\hline 
$0.3-8.0\rm\, keV$ 	   &     40.259 $\pm$      0.014 &     -1.505 $\pm$      0.016 &     -0.421 $\pm$      0.025 &      0.425 $\pm$      0.009 &      0.135 $\pm$      0.009 & &  1.3741 \\
$0.5-2.0\rm\, keV$ 	   &     39.455 $\pm$      0.014 &     -1.514 $\pm$      0.016 &     -0.455 $\pm$      0.025 &      0.433 $\pm$      0.009 &      0.145 $\pm$      0.009 & &  1.4254 \\
$2.0-10.0\rm\, keV$	   &     40.276 $\pm$      0.014 &     -1.503 $\pm$      0.016 &     -0.423 $\pm$      0.025 &      0.425 $\pm$      0.009 &      0.136 $\pm$      0.009 & &  1.3821 \\
$10-100\rm\, keV$  	   &     40.717 $\pm$      0.016 &     -1.417 $\pm$      0.018 &     -0.369 $\pm$      0.027 &      0.394 $\pm$      0.010 &      0.111 $\pm$      0.009 & &  1.6787 \\

\hline \hline
\multicolumn{8}{c}{Based on intrinsic (unabsorbed) SED} \\
\hline 
\colhead{} & 
     \colhead{$\alpha_0$} & 
     \colhead{$\alpha_1$} & 
     \colhead{$\alpha_2$} & 
     \colhead{$\alpha_3$} & 
     \colhead{$\alpha_4$} & 
     \colhead{$\alpha_5$} & 
     \colhead{$\chi^2/N$}\\
	\hline 

$0.3-8.0\rm\, keV$ 	 	& 37.96 $\pm$  0.01 &  0.86 $\pm$  0.02 & -0.213 $\pm$  0.008 &  0.0211 $\pm$  0.0012 & -0.00100 $\pm$  0.00007 &  0.000018 $\pm$  0.000002 &  0.09 \\
$0.5-2.0\rm\, keV$ 	 	& 37.36 $\pm$  0.01 &  0.87 $\pm$  0.02 & -0.214 $\pm$  0.008 &  0.0211 $\pm$  0.0012 & -0.00100 $\pm$  0.00007 &  0.000018 $\pm$  0.000001 &  0.09 \\
$2.0-10.0\rm\, keV$	 	& 37.89 $\pm$  0.01 &  0.86 $\pm$  0.02 & -0.214 $\pm$  0.008 &  0.0213 $\pm$  0.0012 & -0.00101 $\pm$  0.00007 &  0.000018 $\pm$  0.000002 &  0.10 \\
$10-100\rm\, keV$  	 	& 38.38 $\pm$  0.02 &  0.73 $\pm$  0.02 & -0.190 $\pm$  0.010 &  0.0190 $\pm$  0.0014 & -0.00091 $\pm$  0.00009 &  0.000016 $\pm$  0.000002 &  0.14 \\

\hline 
\colhead{} & 
\colhead{$\beta_0$} & 
\colhead{$\beta_1$} & 
\colhead{$\beta_2$} & 
\colhead{$\beta_3$} & 
\colhead{$\beta_4$} & 
\colhead{} & 
\colhead{} \\ 
\hline 

$0.3-8.0\rm\, keV$ 	    &    40.43 $\pm$     0.02 &   -62.39 $\pm$     1.32 &   571.82 $\pm$    13.71 & -1841.42 $\pm$    52.14 &  1976.47 $\pm$    66.28 &  & 0.1751 \\
$0.5-2.0\rm\, keV$ 	    &    39.87 $\pm$     0.02 &   -62.48 $\pm$     1.32 &   572.71 $\pm$    13.71 & -1844.27 $\pm$    52.15 &  1979.52 $\pm$    66.29 &  & 0.1752 \\
$2.0-10.0\rm\, keV$	    &    40.33 $\pm$     0.02 &   -62.20 $\pm$     1.32 &   570.19 $\pm$    13.70 & -1836.17 $\pm$    52.09 &  1970.86 $\pm$    66.21 &  & 0.1748 \\
$10-100\rm\, keV$  	    &    40.54 $\pm$     0.02 &   -61.34 $\pm$     1.34 &   562.26 $\pm$    13.88 & -1810.59 $\pm$    52.81 &  1943.31 $\pm$    67.13 &  & 0.1797 \\

\hline 
\colhead{} & 
\colhead{$\gamma_0$} & 
\colhead{$\gamma_1$} & 
\colhead{$\gamma_2$} & 
\colhead{$\gamma_3$} & 
\colhead{$\gamma_4$} & 
\colhead{} & 
\colhead{} \\ 
\hline 

$0.3-8.0\rm\, keV$ 	  &     40.370 $\pm$      0.016 &     -1.581 $\pm$      0.018 &     -0.495 $\pm$      0.028 &      0.446 $\pm$      0.010 &      0.157 $\pm$      0.010 &  &      1.774\\
$0.5-2.0\rm\, keV$ 	  &     39.795 $\pm$      0.022 &     -1.746 $\pm$      0.025 &     -0.669 $\pm$      0.037 &      0.496 $\pm$      0.014 &      0.207 $\pm$      0.013 &  &      3.199\\
$2.0-10.0\rm\, keV$	  &     40.308 $\pm$      0.015 &     -1.525 $\pm$      0.017 &     -0.445 $\pm$      0.026 &      0.431 $\pm$      0.009 &      0.142 $\pm$      0.009 &  &      1.482\\
$10-100\rm\, keV$  	  &     40.716 $\pm$      0.016 &     -1.418 $\pm$      0.018 &     -0.370 $\pm$      0.027 &      0.394 $\pm$      0.010 &      0.112 $\pm$      0.009 &  &      1.652\\

\enddata
\end{deluxetable*}

\section{DISCUSSION}

Detailed binary population synthesis simulations are used for the first time in order to study the energy feedback of XRBs to the IGM, from $z\sim 20$ until today. Our synthetic XRB models capture accurately all the important physical processes that are involved in the evolution of an XRB population, and have been calibrated to observed XRB population in the local Universe. At the same time these synthetic models are in excellent agreement with X-ray observation of distant normal galaxies up to $z\sim 4$, thus providing a robust framework to study the evolution of XRB populations across cosmic time.

We find that the energetic X-ray photons emitted from XRBs dominate the X-ray radiation field over AGN at $z\gtrsim 6-8$, and hence XRB feedback can be a non-negligible contributor to the heating and re-ionization of the IGM in the early Universe. The SED shape of the XRB emission does not change significantly with redshift, suggesting that the same XRB subpopulation, namely BH XRBs in the high-soft state, dominates the cumulative emission at all time. On the contrary, the normalization of the SED does evolve with redshift. To zeroth order this evolution is driven by the cosmic SFR evolution. However, the metallicity evolution of the Universe and the mean stellar population age are two important factors that affect the X-ray emission from HMXBs and LMXBs, respectively (see Figure~2).

The qualitative effects of an arbitrary X-ray radiation field in the formation and evolution of galaxies have already been studied. \citet{Hambrick2009} performed galaxy evolution simulations incorporating an ultraviolet and X-ray background field. They found that the gas properties at late times are significantly affected by the X-ray component resulting in a 30\% increase of the warm gas component and a four-fold increase in the hot-dense gas component, while at the same time the formation of stars in small systems is reduced. Monte-Carlo realizations of the merger and growth history of BHs show that X-rays from the earliest accreting BHs can provide such a feedback mechanism, on a global scale, finding that the first miniquasars globally warm the intergalactic medium and suppress the formation and growth of subsequent generations of BHs \citep{TPH2012}. More recently, \citet{MFS2013} ran semi-numerical simulations of the dark ages and the epoch of reionization, including both X-rays and ultraviolet radiation fields. They found that X-rays emitted from a XRB population can result in a more extended epoch of reionization and an  overall more uniform reionization morphology, with the largest impact of X-rays being to govern the timing and duration of IGM heating.

All of the aforementioned studies consider either an arbitrary X-ray background field or one that is a mere extrapolation of observations from the local Universe. In this letter we provide analytic prescriptions for the energy feedback from XRBs, based on our detailed synthetic models, which can be directly included in cosmological and galaxy formation and evolution simulations. These new prescriptions allow for the first time not only the qualitative but also the quantitative study of the effects of the energy feedback from XRBs in the early Universe.

\acknowledgements
TF acknowledges support from the CfA and the ITC prize fellowship programs. 
SN is supported by NASA through an Einstein Postdoctoral Fellowship (contract PF2-130096).


\begin{thebibliography}{46}
\expandafter\ifx\csname natexlab\endcsname\relax\def\natexlab#1{#1}\fi

\bibitem[{Aird {et~al.}(2010)Aird, Nandra, Laird, Georgakakis, Ashby, Barmby,
  Coil, Huang, Koekemoer, Steidel, \& Willmer}]{Aird2010}
Aird, J., Nandra, K., Laird, E.~S., Georgakakis, A., Ashby, M. L.~N., Barmby,
  P., Coil, A.~L., Huang, J.~S., Koekemoer, A.~M., Steidel, C.~C., \& Willmer,
  C. N.~A. 2010, \mnras, 401, 2531

\bibitem[{Basu-Zych {et~al.}(2013{\natexlab{a}})Basu-Zych, Lehmer,
  Hornschemeier, Bouwens, Fragos, Oesch, Belczynski, Brandt, Kalogera, Luo,
  Miller, Mullaney, Tzanavaris, Xue, \& Zezas}]{BasuZych2013}
Basu-Zych, A.~R., Lehmer, B.~D., Hornschemeier, A.~E., Bouwens, R.~J., Fragos,
  T., Oesch, P.~A., Belczynski, K., Brandt, W.~N., Kalogera, V., Luo, B.,
  Miller, N., Mullaney, J.~R., Tzanavaris, P., Xue, Y., \& Zezas, A.
  2013{\natexlab{a}}, The Astrophysical Journal, 762, 45

\bibitem[{Basu-Zych {et~al.}(2013{\natexlab{b}})Basu-Zych, Lehmer,
  Hornschemeier, Gon{\c c}alves, Fragos, Heckman, Overzier, Ptak, \&
  Schiminovich}]{BasuZych2013b}
Basu-Zych, A.~R., Lehmer, B.~D., Hornschemeier, A.~E., Gon{\c c}alves, T.~S.,
  Fragos, T., Heckman, T.~M., Overzier, R.~A., Ptak, A.~F., \& Schiminovich, D.
  2013{\natexlab{b}}, The Astrophysical Journal, 774, 152

\bibitem[{Belczynski {et~al.}(2002)Belczynski, kalogera, \&
  Bulik}]{Belczynski2002}
Belczynski, K., kalogera, V., \& Bulik, T. 2002, \apj, 572, 407

\bibitem[{Belczynski {et~al.}(2008)Belczynski, Kalogera, Rasio, Taam, Zezas,
  Bulik, Maccarone, \& Ivanova}]{Belczynski2008}
Belczynski, K., Kalogera, V., Rasio, F.~A., Taam, R.~E., Zezas, A., Bulik, T.,
  Maccarone, T.~J., \& Ivanova, N. 2008, \apjs, 174, 223

\bibitem[{Boroson {et~al.}(2011)Boroson, Kim, \& Fabbiano}]{Boroson2011}
Boroson, B., Kim, D.-W., \& Fabbiano, G. 2011, The Astrophysical Journal, 729,
  12

\bibitem[{Boylan-Kolchin {et~al.}(2009)Boylan-Kolchin, Springel, White,
  Jenkins, \& Lemson}]{Boylan2009}
Boylan-Kolchin, M., Springel, V., White, S.~D.~M., Jenkins, A., \& Lemson, G.
  2009, \mnras, 398, 1150

\bibitem[{Croton {et~al.}(2006)Croton, Springel, White, De~Lucia, Frenk, Gao,
  Jenkins, Kauffmann, Navarro, \& Yoshida}]{Croton2006}
Croton, D.~J., Springel, V., White, S. D.~M., De~Lucia, G., Frenk, C.~S., Gao,
  L., Jenkins, A., Kauffmann, G., Navarro, J.~F., \& Yoshida, N. 2006, \mnras,
  365, 11

\bibitem[{Dib(2011)}]{Dib2011}
Dib, S. 2011, \apjl, 737, L20

\bibitem[{Dijkstra {et~al.}(2012)Dijkstra, Gilfanov, Loeb, \&
  Sunyaev}]{Dijkstra2012}
Dijkstra, M., Gilfanov, M., Loeb, A., \& Sunyaev, R. 2012, \mnras, 421, 213

\bibitem[{Fabbiano(2006)}]{Fabbiano2006}
Fabbiano, G. 2006, \araa, 44, 323

\bibitem[{Faucher-Gigu{\`e}re {et~al.}(2013)Faucher-Gigu{\`e}re, Quataert, \&
  Hopkins}]{FGQH2013}
Faucher-Gigu{\`e}re, C.-A., Quataert, E., \& Hopkins, P.~F. 2013, \mnras, 433,
  1970

\bibitem[{Fragos {et~al.}(2008)Fragos, kalogera, Belczynski, Fabbiano, Kim,
  Brassington, Angelini, Davies, Gallagher, King, Pellegrini, Trinchieri, Zepf,
  Kundu, \& Zezas}]{Fragos2008}
Fragos, T., kalogera, V., Belczynski, K., Fabbiano, G., Kim, D.-W.,
  Brassington, N.~J., Angelini, L., Davies, R.~L., Gallagher, J.~S., King,
  A.~R., Pellegrini, S., Trinchieri, G., Zepf, S.~E., Kundu, A., \& Zezas, A.
  2008, The Astrophysical Journal, 683, 346

\bibitem[{Fragos {et~al.}(2009)Fragos, kalogera, Willems, Belczynski, Fabbiano,
  Brassington, Kim, Angelini, Davies, Gallagher, King, Pellegrini, Trinchieri,
  Zepf, \& Zezas}]{Fragos2009}
Fragos, T., kalogera, V., Willems, B., Belczynski, K., Fabbiano, G.,
  Brassington, N.~J., Kim, D.-W., Angelini, L., Davies, R.~L., Gallagher,
  J.~S., King, A.~R., Pellegrini, S., Trinchieri, G., Zepf, S.~E., \& Zezas, A.
  2009, \apjl, 702, L143

\bibitem[{Fragos {et~al.}(2013)Fragos, Lehmer, Tremmel, Tzanavaris, Basu-Zych,
  Belczynski, Hornschemeier, JENKINS, kalogera, Ptak, \& Zezas}]{Fragos2013}
Fragos, T., Lehmer, B., Tremmel, M., Tzanavaris, P., Basu-Zych, A., Belczynski,
  K., Hornschemeier, A., JENKINS, L., kalogera, V., Ptak, A., \& Zezas, A.
  2013, The Astrophysical Journal, 764, 41

\bibitem[{Guo {et~al.}(2011)Guo, White, Boylan-Kolchin, De~Lucia, Kauffmann,
  Lemson, Li, Springel, \& Weinmann}]{Guo2011}
Guo, Q., White, S., Boylan-Kolchin, M., De~Lucia, G., Kauffmann, G., Lemson,
  G., Li, C., Springel, V., \& Weinmann, S. 2011, \mnras, 413, 101

\bibitem[{Haiman(2011)}]{Zoltan2011}
Haiman, Z. 2011, Nature, 472, 47

\bibitem[{Hambrick {et~al.}(2009)Hambrick, Ostriker, Naab, \&
  Johansson}]{Hambrick2009}
Hambrick, D.~C., Ostriker, J.~P., Naab, T., \& Johansson, P.~H. 2009, The
  Astrophysical Journal, 705, 1566

\bibitem[{Hopkins {et~al.}(2011)Hopkins, Quataert, \& Murray}]{Hopkins2011}
Hopkins, P.~F., Quataert, E., \& Murray, N. 2011, \mnras, 417, 950

\bibitem[{Hopkins {et~al.}(2007)Hopkins, Richards, \& Hernquist}]{HRH2007}
Hopkins, P.~F., Richards, G.~T., \& Hernquist, L. 2007, The Astrophysical
  Journal, 654, 731

\bibitem[{Justham \& Schawinski(2012)}]{JS2012}
Justham, S. \& Schawinski, K. 2012, \mnras, 423, 1641

\bibitem[{Lehmer {et~al.}(2010)Lehmer, Alexander, Bauer, Brandt, Goulding,
  {Jenkins, L. P.}, Ptak, \& Roberts}]{Lehmer2010}
Lehmer, B.~D., Alexander, D.~M., Bauer, F.~E., Brandt, W.~N., Goulding, A.~D.,
  {Jenkins, L. P.}, Ptak, A., \& Roberts, T.~P. 2010, The Astrophysical
  Journal, 724, 559

\bibitem[{Lehmer {et~al.}(2012)Lehmer, Xue, Brandt, Alexander, Bauer, Brusa,
  Comastri, Gilli, Hornschemeier, Luo, Paolillo, Ptak, Shemmer, Schneider,
  Tozzi, \& Vignali}]{Lehmer2012}
Lehmer, B.~D., Xue, Y.~Q., Brandt, W.~N., Alexander, D.~M., Bauer, F.~E.,
  Brusa, M., Comastri, A., Gilli, R., Hornschemeier, A.~E., Luo, B., Paolillo,
  M., Ptak, A., Shemmer, O., Schneider, D.~P., Tozzi, P., \& Vignali, C. 2012,
  The Astrophysical Journal, 752, 46

\bibitem[{Loeb \& Barkana(2001)}]{Loeb2001}
Loeb, A. \& Barkana, R. 2001, \araa, 39, 19

\bibitem[{McClintock \& Remillard(2006)}]{McCR2006}
McClintock, J.~E. \& Remillard, R.~A. 2006, In: Compact stellar X-ray sources.
  Edited by Walter Lewin {\&} Michiel van der Klis. Cambridge Astrophysics
  Series, 157

\bibitem[{McQuinn {et~al.}(2007{\natexlab{a}})McQuinn, Hernquist, Zaldarriaga,
  \& Dutta}]{McQ2007}
McQuinn, M., Hernquist, L., Zaldarriaga, M., \& Dutta, S. 2007{\natexlab{a}},
  \mnras, 381, 75

\bibitem[{McQuinn {et~al.}(2007{\natexlab{b}})McQuinn, Lidz, Zahn, Dutta,
  Hernquist, \& Zaldarriaga}]{McQ2007b}
McQuinn, M., Lidz, A., Zahn, O., Dutta, S., Hernquist, L., \& Zaldarriaga, M.
  2007{\natexlab{b}}, \mnras, 377, 1043

\bibitem[{Mesinger {et~al.}(2013)Mesinger, Ferrara, \& Spiegel}]{MFS2013}
Mesinger, A., Ferrara, A., \& Spiegel, D.~S. 2013, \mnras, 431, 621

\bibitem[{Mineo {et~al.}(2012{\natexlab{a}})Mineo, Gilfanov, \&
  Sunyaev}]{Mineo2012a}
Mineo, S., Gilfanov, M., \& Sunyaev, R. 2012{\natexlab{a}}, \mnras, 419, 2095

\bibitem[{Mineo {et~al.}(2012{\natexlab{b}})Mineo, Gilfanov, \&
  Sunyaev}]{Mineo2012b}
---. 2012{\natexlab{b}}, arXiv, 2157

\bibitem[{Mirabel {et~al.}(2011)Mirabel, Dijkstra, Laurent, Loeb, \&
  Pritchard}]{Mirabel2011}
Mirabel, I.~F., Dijkstra, M., Laurent, P., Loeb, A., \& Pritchard, J.~R. 2011,
  \aap, 528, A149+

\bibitem[{Ostriker \& Gnedin(1996)}]{OG1996}
Ostriker, J.~P. \& Gnedin, N.~Y. 1996, Astrophysical Journal Letters v.472,
  472, L63

\bibitem[{Power {et~al.}(2012)Power, James, Combet, \& Wynn}]{Power2012}
Power, C., James, G.~F., Combet, C., \& Wynn, G. 2012, arXiv, 5854

\bibitem[{Pritchard \& Furlanetto(2007)}]{PF2007}
Pritchard, J.~R. \& Furlanetto, S.~R. 2007, \mnras, 376, 1680

\bibitem[{Sales {et~al.}(2010)Sales, Navarro, Schaye, Dalla~Vecchia, Springel,
  \& Booth}]{Sales2010}
Sales, L.~V., Navarro, J.~F., Schaye, J., Dalla~Vecchia, C., Springel, V., \&
  Booth, C.~M. 2010, \mnras, 409, 1541

\bibitem[{Silverman {et~al.}(2008)Silverman, Green, Barkhouse, Kim, Kim,
  Wilkes, Cameron, Hasinger, Jannuzi, Smith, Smith, \&
  Tananbaum}]{Silverman2008}
Silverman, J.~D., Green, P.~J., Barkhouse, W.~A., Kim, D.-W., Kim, M., Wilkes,
  B.~J., Cameron, R.~A., Hasinger, G., Jannuzi, B.~T., Smith, M.~G., Smith,
  P.~S., \& Tananbaum, H. 2008, The Astrophysical Journal, 679, 118

\bibitem[{Sobacchi \& Mesinger(2013)}]{SM2013}
Sobacchi, E. \& Mesinger, A. 2013, \mnras, 432, 3340

\bibitem[{Sobolewska \& {\.Z}ycki(2006)}]{SZ2006}
Sobolewska, M.~A. \& {\.Z}ycki, P.~T. 2006, \mnras, 370, 405

\bibitem[{Springel {et~al.}(2005)Springel, Di~Matteo, \& Hernquist}]{SMH2005}
Springel, V., Di~Matteo, T., \& Hernquist, L. 2005, \mnras, 361, 776

\bibitem[{Stinson {et~al.}(2013)Stinson, Brook, Macci{\`o}, Wadsley, Quinn, \&
  Couchman}]{Stinson2013}
Stinson, G.~S., Brook, C., Macci{\`o}, A.~V., Wadsley, J., Quinn, T.~R., \&
  Couchman, H. M.~P. 2013, \mnras, 428, 129

\bibitem[{Tanaka {et~al.}(2012)Tanaka, Perna, \& Haiman}]{TPH2012}
Tanaka, T., Perna, R., \& Haiman, Z. 2012, \mnras, 425, 2974

\bibitem[{Tremmel {et~al.}(2013)Tremmel, Fragos, Lehmer, Tzanavaris,
  Belczynski, kalogera, Basu-Zych, Farr, Hornschemeier, JENKINS, Ptak, \&
  Zezas}]{Tremmel2013}
Tremmel, M., Fragos, T., Lehmer, B.~D., Tzanavaris, P., Belczynski, K.,
  kalogera, V., Basu-Zych, A.~R., Farr, W.~M., Hornschemeier, A., JENKINS, L.,
  Ptak, A., \& Zezas, A. 2013, The Astrophysical Journal, 766, 19

\bibitem[{Tzanavaris \& Georgantopoulos(2008)}]{TG2008}
Tzanavaris, P. \& Georgantopoulos, I. 2008, \aap, 480, 663

\bibitem[{van Dokkum(2008)}]{Dokkum2008}
van Dokkum, P.~G. 2008, The Astrophysical Journal, 674, 29

\bibitem[{Wu {et~al.}(2010)Wu, Yu, Li, Maccarone, \& Li}]{Wu2010}
Wu, Y.~X., Yu, W., Li, T.~P., Maccarone, T.~J., \& Li, X.~D. 2010, The
  Astrophysical Journal, 718, 620

\bibitem[{Zhang {et~al.}(2012)Zhang, Gilfanov, \& Bogd{\'a}n}]{ZGA2012}
Zhang, Z., Gilfanov, M., \& Bogd{\'a}n, {\'A}. 2012, \aap, 546, 36

\end{thebibliography}

\end{document}